\documentclass[aps,pra,onecolumn,groupedaddress,amsfonts,amssymb]{revtex4}
\usepackage{graphics}
\usepackage{epsfig}
\usepackage{amsmath}
\usepackage{amsthm}

\def\comment#1{}

\begin{document}

\title{Using Quantum Mechanics to Cope with Liars}

\author{Stefano Mancini}
\email{stefano.mancini@unicam.it}
\affiliation{Dipartimento di Fisica, Universit\`{a} di Camerino, 
I-62032 Camerino, Italy\\
\& INFN Sezione di Perugia, I-06123 Perugia, Italy}
\author{Lorenzo Maccone}\email{maccone@quantummechanics.it} 
\affiliation{QUIT - Quantum Information Theory Group, Dipartimento di Fisica
  ``A.  Volta'' Universit\`a di Pavia, via A.  Bassi 6, I-27100 Pavia,
  Italy.}

\date{\today}

\begin{abstract}
We propose the use of a quantum algorithm to deal with the problem of searching with errors in the framework of two-person games.
Specifically, we present a solution to the Ulam's problem 
that polynomially reduces its query complexity  
and makes it independent from the dimension of the search space.
\end{abstract}
\maketitle

\section{Introduction}
In 1976 S. Ulam\cite{Ulam} raised the following question, which
subsequently became known as the ``Ulam problem''\footnote{It is also
  known as ``R\'enyi-Ulam problem'', because a similar game was first
  proposed by R\'enyi in 1961\cite{Renyi}.}:

\bigskip
\textit{Someone thinks of a number between one and one million (which
  is just less than $2^{20}$).  Another person is allowed to ask up to
  twenty questions, to each of which the first person is supposed to
  answer only yes or no.  Obviously the number can be guessed by
  asking first: Is the number in the first half-million? and then
  again reduce the reservoir of numbers in the next question by
  one-half, and so on. Finally the number is obtained in less than
  $\log_2(1,000,000)$. Now suppose one were allowed to lie once or
  twice, then how many questions would only need to get the right
  answer? One clearly needs more than $n$ questions for guessing one
  of the $2^n$ objects because one does not know when the lie was
  told. This problem is not solved in general.}
\bigskip

One can consider the Ulam problem as an interactive game between two
players, Alice (the Questioner) and Bob (the Responder). 
Bob thinks of a number $a$ in the set
${\cal S}\equiv\{1\ldots N\}$ (hereafter we assume $N=2^n$ for the sake of
simplicity) and Alice has to find the number $a$ by asking yes-no
queries of the type ``$a\in S$?'', where $S$ is any subset of ${\cal S}$. 
The game is played interactively, i.e. each query is
answered before the next query is stated. The solution to the Ulam
problem is the minimum number $k_*$ of such yes-no queries required to
find the number $a$, provided Bob may lie $l$ times.

It is a fully adaptive binary search with arbitrary questions and a
fixed upper bound on the number of lies. It corresponds to a
communication through a noisy channel with noiseless feedback, where
we assume that at most $l$ errors can be made during the entire
transmission.  As an oracle problem, a lower bound of $\Omega(n+l\log
n)$ oracle queries has been established in Ref.~\cite{Rivest}.

The speedup of quantum algorithms over classical algorithms is the
main reason for the current interest on quantum computing.  However,
if we simply translate the Ulam problem to a quantum search, then a
lower bound $\Omega(n)$ for ordered searching\cite{OS} is encountered
already for $l=0$.  Hence, the problem seems to admit no quantum
speedup.

Nevertheless, going beyond quantum search, we shall provide a more 
efficient solution.

\section{The Classical Pathway}\label{s:classical}

The volume conservation law\cite{Berlekamp} indicates the method to
produce the shortest possible strategy for Alice. In every state of
the game, she should ask a question that splits the volume of the
state as evenly as possible.

The volume bound turns out to be equal to the Hamming sphere-packing
bound\cite{sloane}, i.e. $2^k\ge\sum_{i=0}^{l}\left(\begin{array}{c}k\\
i\end{array}\right) 2^n$.  Such a bound is known to be achievable for
$l=0,1,2,3$ (see e.g. Ref.~\cite{Pelc} for a review on the results to
the Ulam problem).  Hence, the minimum number $k_*$ of yes-no queries
(the solution to the Ulam problem) is the lowest integer $k$ satisfying
the inequalities:
\begin{description}
\item{i) $k\ge n$ for $l=0$ (which gives $k_*=20$ for $n=20$).}\\
\item{ii) $\frac{2^k}{k+1}\ge2^n$ for $l=1$ (which gives $k_*=25$     for
$n=20$).}\\
\item{iii) $\frac{2^{k+1}}{k^2+k+2}\ge2^n$ for $l=2$ (which gives $k_*=29$
    for $n=20$).}\\
\item{iv) $3\frac{2^{k+1}}{k^3+5k+1}\ge2^n$ for $l=3$     (which gives
$k_*=33$ for $n=20$).}
\end{description}
For larger values of $l$, exact results for the minimum length of
Alice's strategy are valid only for particular values of the search
space dimension $N$\cite{Pelc}.

\medskip
 
For our purposes we are now going to consider a simple non-adaptive strategy starting from the case of  no lies.

Say $B\equiv\{0,1\}$ the usual binary field, then $a$ would be a
vector in $B^n$ since in this case $B^n\equiv{\cal S}$.  Note that for
Alice each string $x\in B^n$ defines two subspaces of the search
space: that of even parity $\{x\in B^n|a\cdot x=0\}$, and that of odd
parity $\{x\in B^n|a\cdot x=1\}$.  Here ``$\cdot$'' stands for the
usual scalar product in $B^n$, i.e. $a \cdot x \equiv$mod$_2(\sum_j a_j
x_j)$, where $a_j$ and $x_j$ are the $j$th bits of $a$ and $x$
respectively.  Let us define $S_x$ the former subspace, then the
question ``$a\in S_x$?'' translates into the evaluation of $a\cdot x$,
with the convention that the value $0$ is equivalent to ``YES'' answer
and the value $1$ is equivalent to ``NO'' answer.  Thus, the game
corresponds to the evaluation of the function $f_a:B^n\to B$
parametrized by $a$ and such that $f_a(x)=a\cdot x$.

To find $a$ Alice does not need to use all possible $x\in B^n$, but it
is sufficient for her to pose questions using inputs $x$ with only
the $k$th bit equal to $1$ (and all the rest equal to $0$). This is
equivalent to asking questions of the type ``Is the $k$th bit equal to
$1$?''; $n$ of these questions will allow her to obtain the value of
$a$.

\section{A Quantum Shortcut}

Since the goal is to determine the parameter $a$ from the above
function evaluation, the problem resembles that of
Bernstein-Vazirani\cite{Vaz}, and one can resort to the algorithm
originally proposed by Deutsch and Jozsa\cite{DJ}.  It requires Alice
to pose an equally weighted superposition of all possible questions,
which can be achieved through Hadamard operations: On an $n$-qubits
register, these are given by   $H^{\otimes
n}=\frac{1}{\sqrt{2^n}}\sum_{x,y\in B^n}(-1)^{x\cdot
y}|x\rangle\langle y|$. Then, a clever use of quantum interference
allows her to recover the parameter $a$, from a single query to Bob.
Two quantum registers are necessary: A register ${\cal Q}$ composed of
$n$ qubits where Alice stores the input $x$, and a register ${\cal R}$
composed by one qubit where Bob stores the answer. When prompted with
$|x\rangle_{\cal Q}\:|y\rangle_{\cal R}$, he returns   $|x\rangle_{\cal
Q}\:|y\oplus a\cdot x\rangle_{\cal R}$.  In
detail, the algorithm is composed of the following
steps:

\begin{itemize}
\item{ Alice initializes the registers ${\cal Q}$ and ${\cal R}$     as
$|0\rangle_{\cal Q}\;|1\rangle_{\cal R}$,}\\
\item{and applies a Hadamard transform to both registers ${\cal Q}$ and 
${\cal R}$:
\begin{equation}
\frac{1}{\sqrt{2^n}}\sum_{x\in B^n}|x\rangle_{\cal Q}\;
\frac{1}{\sqrt{2}}\left(|0\rangle_{\cal R}-|1\rangle_{\cal R}\right).
\end{equation}
}
\item{Bob then evaluates the function $f_a$:
\begin{equation}
\frac{1}{\sqrt{2^n}}\sum_{x\in B^n}(-1)^{a\cdot x}|x\rangle_{\cal Q}\;
\frac{1}{\sqrt{2}}\left(|0\rangle_{\cal R}-|1\rangle_{\cal R}\right).
\end{equation}
}
\item{Alice applies a Hadamard transform to the register ${\cal Q}$:
\begin{equation}\label{four}
\frac{1}{2^n}\sum_{y\in B^n}\left[\sum_{x\in B^n}(-1)^{a\cdot x\oplus 
y\cdot x}\right]|y\rangle_{\cal 
Q}\;\frac{1}{\sqrt{2}}\left(|0\rangle_{\cal R}-|1\rangle_{\cal 
R}\right),
\end{equation}
}
\item{and measures the register ${\cal Q}$ in the computational basis.
  }
\end{itemize}
Note that the only amplitude different from zero in the term inside
the square brackets of Eq.~(\ref{four}) is the one for which $a\oplus
y=0$. This implies that $y=a$ is the only possible result to the 
measurement.  This is a consequence of the fact that
\begin{equation}
\sum_{x\in B^n}(-1)^{a\cdot x\oplus y\cdot x}=2^n\delta_{a,y}\,.
\end{equation}
Thus, if $l=0$, it is possible to determine $a$ with only one query
instead of the $\Omega(n)$ that are necessary to any classical
strategy.

If Bob is allowed to lie, then he may choose a parameter $a'$
different from $a$, and Alice will not be able to recover the true
value $a$ from the above protocol. However, since he cannot lie more
than $l$ times, she can repeat the protocol $2l+1$ times, and then
apply a majority-voting strategy\footnote{This strategy works also
  when the number of lies is not exactly known, but $l$ only
  represents an upper bound on it.}: Of the $2l+1$ answers she
obtained from the repetition of the protocol, no less than $l+1$ will
be coincident, and they will be all equal to the correct parameter
$a$. This gives $k_*=1$, $3$, $5$, $7$ respectively for the examples i), ii), iii),
iv) of Sec. 2.

Furthermore, we may distinguish the number of lies from the number of
bits Bob flips in passing from $a$ to $a'$ (for each lie he can flip
one or more bits of $a$).  If there is a constraint $l^*$ on the
maximum number of bits Bob can flip on the whole game, then a more
efficient strategy can be devised. For instance, for even $l^*$ the
number of queries can be reduced to $l^*+1$, and for odd $l^*$ it can
be reduced to $l^*+2$.

Finally, notice that it does not matter if Bob lies in also by
inverting the value of the function evaluation, i.e. by returning
$f_a(x)=a\cdot x\oplus 1$ instead of $f_a(x)=a\cdot x$.  In fact, in
such a case the term inside the square brackets of Eq.(\ref{four})
becomes
\begin{equation}
\sum_{x\in B^n}(-1)^{a\cdot x\oplus y\cdot x\oplus 
1}=-2^n\delta_{a,y}\,.
\end{equation}
As before, it is always equal to zero except when $a\oplus y=0$, i.e.
$y=a$ is again the only possible result for Alice's measurement, and
she finds the right answer.

\section{Conclusion}

Summarizing, it is shown that the Ulam problem is exactly solvable in
the quantum framework where its query complexity reduces from
$\Omega(n+l\log n)$ to $O(2l+1)$.  
Whenever $N\neq 2^n$ it suffices to repeat
the above arguments with $n=\lceil \log N\rceil$.

Different questions through which formulate the Ulam game could be revisited in a quantum framework.
Since the game is also viewed as a tool for interpreting some problems in logic\cite{Mundici},
this may have some impact in the field of quantum logic.
It also turns out that the Ulam problem is remarkably similar to the
solution of one of the main problems in coding theory, namely finding
the minimum length for a code of a given size and a given minimum
distance\cite{Pelc}. Hence, it would be interesting to explore possible
implications on quantum error correction\cite{QEC}.





\section*{Acknowledgments}

S. M. is grateful to Richard Jozsa and Daniele Mundici for
enlightening discussions. L. M. thanks the Department of Botany and
Ecology of the University of Camerino for the kind hospitality and acknowledges 
financial support by MIUR through Cofinanziamento 2003 and EC through ATESIT (Contract No. IST-2000-29681).



\end{document}